\documentclass[12pt,oneside,reqno]{amsart}
\usepackage{amssymb}
\usepackage[final]{showkeys}
\usepackage{microtype}
\usepackage{color}
\usepackage{graphicx}
\usepackage[hmargin=3.5cm,vmargin={5cm,5cm}]{geometry}

\newcommand{\ds}{\displaystyle}
\numberwithin{equation}{section}

\allowdisplaybreaks

\def\e{{\rm e}}

\def\Ei{{\rm Ei}\,}
\def\Ein{{\rm Ein}\,}
\def\log{{\rm log}\,}
\def\EE{{\mathcal E}}

 \numberwithin{equation}{section}
\begin{document}

    \title[]{A generalization of the Becker model\\ 
    in linear viscoelasticity: creep, relaxation \\
    and internal friction}

    \author{Francesco Mainardi$^1$}
    	 \address{${}^1$Dipartimento di Fisica e Astronomia (DIFA), 
    	 University of  Bologna ``Alma Mater Studiorum", and INFN, 
    	 Via irnerio 46, I-40126 Bologna, Italy}
	 \email{francesco.mainardi@bo.infn.it  \  (Corresponding Author)}
    
 \author{Enrico  Masina$^2$}
     \address{${}^2$Dipartimento di Fisica e Astronomia (DIFA)
    University of  Bologna  ``Alma Mater Studiorum", and INFN, I-40126 Via Irnerio 46, I-40126 Bologna, Italy.}
    \email{enrico.masina@bo.infn.it}

    \author{Giorgio Spada$^3$}
   \address{${}^3$Dipartimento  di Scienze Pure e Applicate
             (DiSPeA),  University of Urbino ``Carlo Bo", Via Santa Chiara 27,
             I- 61029 Urbino, Italy.} 

\email{giorgio.spada@gmail.com}    


    \date{\today}

    \begin{abstract}
    We present a new rheological model
    depending on a real parameter $\nu \in [0,1]$ that reduces to the Maxwell body for $\nu=0$ and to the Becker body for $\nu=1$.
The corresponding creep law  is expressed in an  integral form in which the exponential 
function of the Becker model is replaced and generalized  by a Mittag-Leffler function of order $\nu$. 
Then,   the corresponding non-dimensional creep function      and its rate
{are studied as functions of   time for different values of $\nu$ in order }to visualize the transition  from the classical
        Maxwell body to the Becker body.
    Based on the hereditary theory of linear viscoelasticity,
     we also {approximate      the  relaxation function} by solving numerically a Volterra integral equation of the second kind.
     In  turn, the relaxation function is shown  
     versus time for different values 
     of $\nu$ to visualize again the transition  from the classical Maxwell body to the Becker body.
    Furthermore,  we provide a full characterization of the new model  by computing, in addition  to the  creep and  relaxation functions, the so-called specific dissipation $Q^{-1}$  as a function  of frequency, which is of  particularly relevance for  geophysical applications.
    
	\smallskip
	\noindent{\bf Published on line in Mechanics of Time-Dependent Materials,
	  on  2 February 2018,  pp 12;
	  DOI:10.1007/s11043-018-9381-4} 
	
    \end{abstract}

    \maketitle

    \section{{Introduction. The Becker  model: the creep law and the spectra}}

    In 1925 Becker\footnote{%
    Richard Becker (1887--1955)  was a German theoretical
physicist who made relevant contributions in thermodynamics, statistical mechanics, electromagnetism,  superconductivity, and quantum electrodynamics. He was professor  formerly in Berlin and then in Gottingen. 
    For more details see $https://en.wikipedia.org/wiki/Richard{\_}Becker{\_}(physicist)$.}
     introduced a creep law to deal with the deformation of particular viscoelastic and plastic bodies \cite{Becker_1925}.
 {Here we aim at extending the Becker law in order to enlarge its possible
spectrum of applications. In this Section, we briefly summarize the creep and the
spectral properties of the Becker rheological law using some basic concepts of linear
viscoelasticity,  illustrated in the Appendix. Subsequently, we generalize the Becker
law and study its creep behavior (Section 2), the rate of creep and the spectra
(Section 3), and the relaxation properties (Section 4). In Section 5 we obtain the
specific dissipation function ($Q^{-1}$) and we study its frequency dependency. Our
conclusions are drawn in Section 6. }    
     
 {The creep law proposed by Becker}
provides  the strain response $\epsilon(t)$ to a constant stress $\sigma(t)=\sigma_0$  in the form
     \begin{equation}\label{lo}
     \epsilon(t)= \frac{\sigma_0}{E_0}\left[1+q\, \Ein (t/\tau_0)\right],
     \quad t\ge 0\,,
     \end{equation}
     where $E_0$ is the shear modulus, $\tau_0>0$  is a characteristic  time during which the transition from 
     elastic to creep-type deformation occurs 
     and $q>0$ is a  non-dimensional constant.    
    The function $\Ein(z)$ is a transcendental function first introduced by Schelkunoff in 1944
    \cite{Schelkunoff_1944}  and defined as
   \begin{equation}
\Ein (t) = \int_0^{z} \frac{1 - \e^{-u}}{u}\,du\,, \quad |\hbox{arg}\, z|<\pi\,,     
\label{Psi}
\end{equation} 
and related to the exponential integral 
$ \EE_1(z)$ and to the incomplete Gamma function
$\Gamma(0,z)= 
{\ds \int_z^\infty\! \frac{\e^{-u}}{u} \,  du} $ 
as
\begin{equation}
\EE_1(z) =   -\Ei(-z)  = \Gamma(0\,,z) =
 -C  - \log z + \Ein (z)\,,
   \end{equation}
with $|\hbox{arg} \,z| < \pi$ and where
$C= -\Gamma^\prime (1)= 0.577215\dots$
 denotes the Euler-Mascheroni constant.
For further mathematical details on the exponential integral and its generalizations we refer the reader to the NIST Handbook \cite{NIST}.
{Additional results  shall be presented 
in  Masina and Mainardi  \cite{Masina-Mainardi_2018}}.

We note that originally Becker was not aware of the $\Ein$ function 
(introduced in 1944) but only of the classical exponential integral.

  The  creep law proposed  by Becker on the basis of empirical arguments  has found a number of applications, 
formerly  in ferromagnetism, see the 1939 treatise by Becker and Doring \cite{Becker-Doring_BOOK39}, and in mathematical theory of linear viscoelasticity, see e.g. 
  Gross (1953) \cite{Gross_BOOK53},
  in which we find references to applications in dielectrics in the 1950's.
  In 1956 Jellinek and Brill \cite{Jellinek-Brill_1956} proposed a model for the primary creep of ice 
  based on the Becker model.   
  In 1967 Orowan \cite{Orowan_1967} recalled the Becker model in order to get a
  $Q$ quality factor for dissipation almost independent on frequency as observed in most rheological materials, mainly in Seismology. Indeed, in view of this weak dependence of the $Q$ factor in Seismology,  in 1982 Strick and Mainardi  
  \cite{Strick-Mainardi_1982} have investigated  the Becker model in comparison with the most famous  Lomnitz model of logarithmic creep.
   {We also recall the papers by Neubert (1963) \cite{Neubert_AQ1963}, by Holenstein et al. (1980)
  \cite{Holenstein-et-al_JBE1980} and the recent one by Hanyga \cite{Hanyga_PAGEOPH2014}, where the Becker model was considered for representing internal damping in solid materials, in arterial viscoelastic walls, and in seismology, respectively.}   
  Unfortunately, in spite of its benefits,  the Becker model was then  neglected in the rheological literature, {but shortly recalled in the 2010 book by Mainardi \cite{Mainardi_BOOK10}}. 
  Nevertheless, in linear viscoelasticity the Becker law was (independently) rediscovered in 1992 by Lubliner and Panoskaltsis as a modification of the 1947 Kuhn { logarithmic creep} law
  \cite{Lubliner-Panoskaltsis_1992}, 
  but the priority of Becker with  respect to  Kuhn is out of discussion.

Herewith, in Fig. 1  we show the plots of the creep function $\psi(t)$  for the original  Becker model with comparison to its asymptotic representations for small and large times, as pointed out in the books on special functions, see e.g.  \cite{NIST}, namely
  \begin{equation}
\psi(t)= \Ein (t) \sim
\left\{
\begin{array}{ll}
t- \frac{t^2}{4} \,, & t \to 0^+\,, \\ \\
\log (t) + C\,,  & t \to +\infty\,.
\end{array}
\right.
\end{equation}
 
    \begin{figure}[h!!]
    \begin{center}
\includegraphics[scale=0.65]
{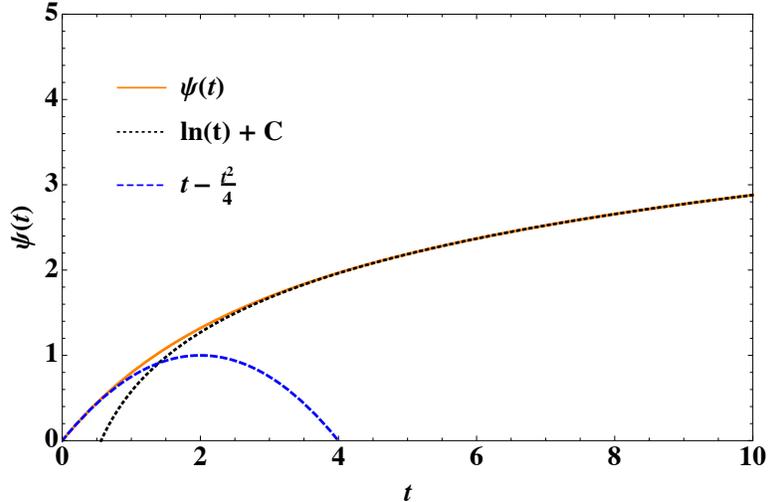}
 \end{center}
 \vskip-0.5truecm
\caption{The creep function $\psi(t)$  for the original Becker model compared with its asymptotic representations given by Eq.(1.4)}
\end{figure}
    
   {The spectra of the Becker model are easily derived   
 from the corresponding Laplace transform of the rate of creep:
 \begin{equation}
 s \tilde \psi(s) = 
\mathcal{L}\left\{\frac{d\psi}{dt}\right \} =
 \ln\left(1 + \frac{1}{s}\right)\,.
 \end{equation}
 Indeed, by using the Titchmarsh formula \eqref{K(r)} and 
 Eq. \eqref{H(tau)},   {we get for the frequency and time spectra:}
 \begin{equation}
 K(r)=
 \left\{
 \begin{array}{ll}  
1 & 0\le r <1, \\
0  & 1\le r<\infty;
\end{array}
\right.
\quad
H(\tau)=
 \left\{
 \begin{array}{ll}  
0 & 0 \le \tau <1, \\
1/\tau^2  & 1\le \tau<\infty,
\end{array}
\right.
\end{equation}
{respectively}.
{The plots of the two spectra are  shown in   Fig. 2.} 
{Since the above spectra are non-negative, the  completely monotonicity (CM) property of the rate of creep follows
from the Bernstein theorem, see e.g. \cite{sch}}.
{However the CM property is already ensured in view a well known theorem according to which a non negative, finite linear combination of CM functions is a CM function, see again \cite{sch}. Indeed,
in view of \eqref{Psi},  the rate of creep 
can be written as}
\begin{equation}
\frac{d \psi}{dt}(t) = \frac{1}{t} - \frac{\e^{-t}}{t}  \in[0,1] \quad t\ge0\,.
 \label{rate-of-creep}
\end{equation}

\begin{figure}
    \centering
    \includegraphics[width=6.5cm]{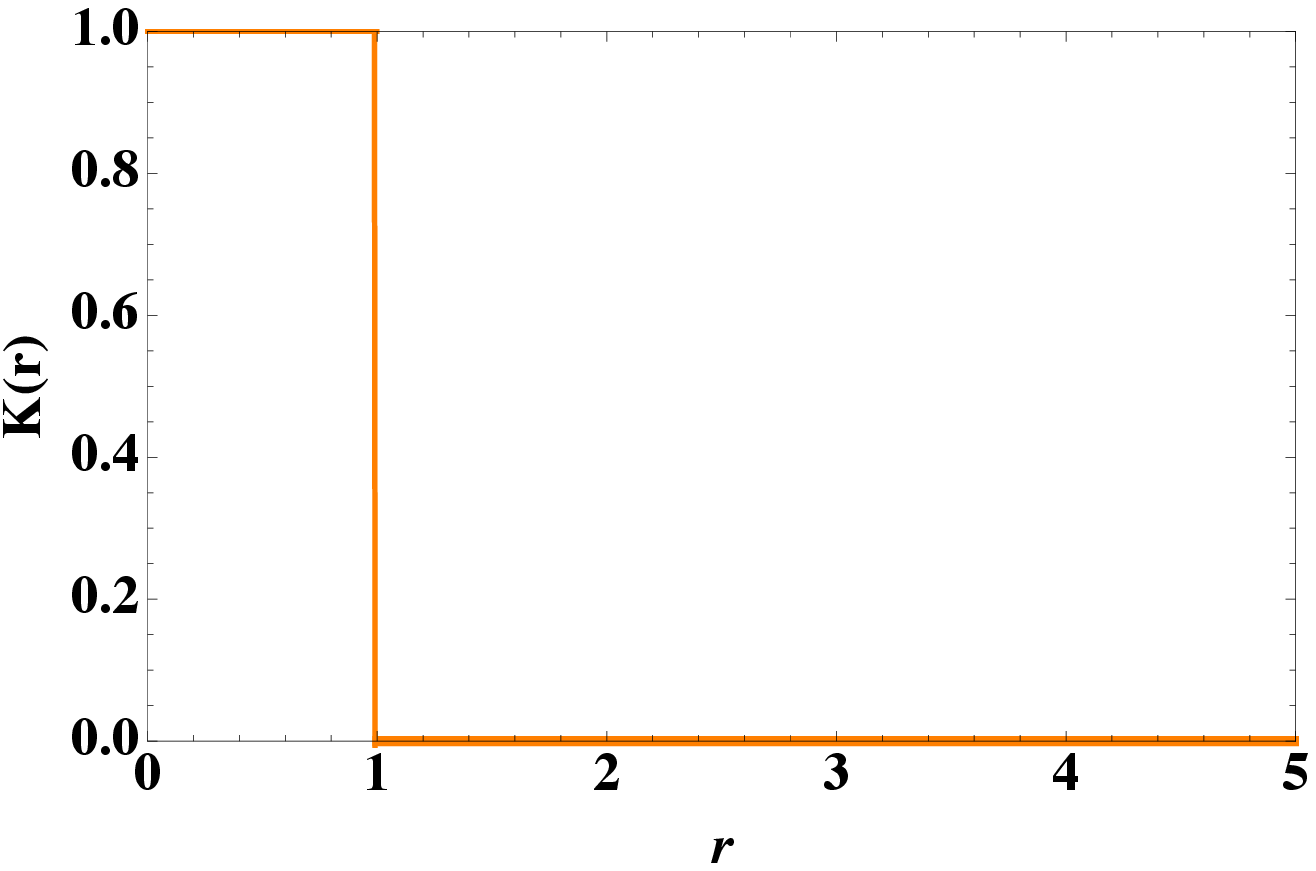}%
    $\quad$
    \includegraphics[width=6.5cm]{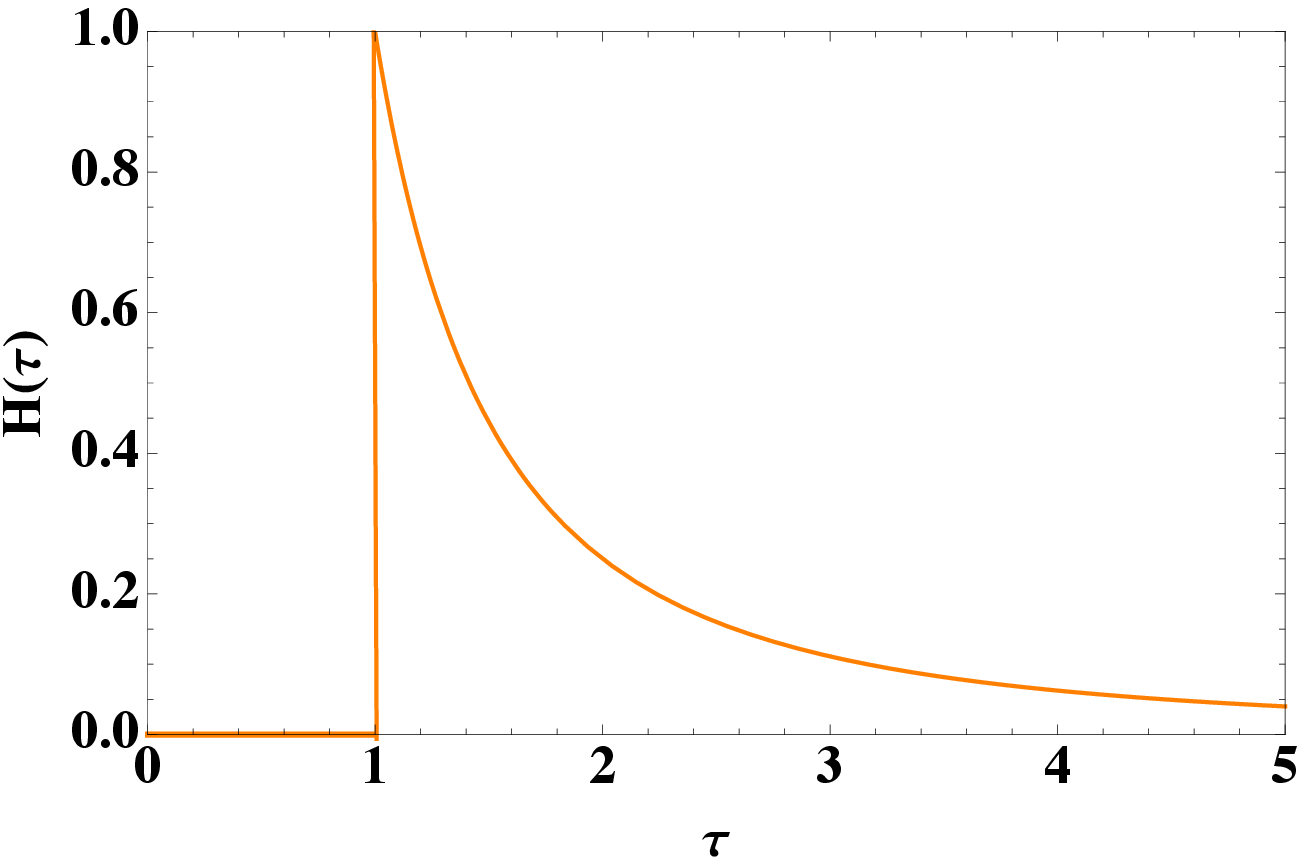} %
     \vskip-0.5truecm
    \caption{The spectra for the Becker body: left in frequency $K(r)$; right in time $H(\tau)$.}%
    \label{fig:Becker-spectra}%
\end{figure}

We  note that in his 1953 treatise, Gross \cite{Gross_BOOK53}
has well pointed out the existence of the spectra for the Becker body without citing the equivalent  CM properties being these mathematical notions unknown to Becker himself  and to  him. 
{We also note that more recently Mainardi and  Spada  (2012) 
\cite{Mainardi-Spada_2012a} have revisited the creep spectra of the Becker body in comparison with those of the Lomnitz logarithmic creep model.}
\section{The generalized Becker model: {the creep function}}

  Let us now consider our generalization of the Becker model 
  by writing the new creep compliance as depending on a real parameter 
  $\nu \in (0,1]$
\begin{equation}
{J}_{\nu}(t) ={J}_0 [1 + q\, \psi_{\nu}(t)]\,,
\end{equation}
where 
\begin{equation}
\psi_{\nu}(t) = \Gamma(\nu + 1)\Ein_{\nu}(t)
\label{Psi-nu}
\end{equation}
with 
\begin{equation}
\Ein_{\nu}(t) = \int_0^{t} \frac{1 - E_{\nu}(-u^{\nu})}{u^{\nu}}\, du\,
\label{Ein-nu}.
\end{equation}
Above we have introduced  the Mittag-Leffler function
\begin{equation}
E_{\nu}(-u^{\nu}) \equiv E_{\nu, 1}(-u^{\nu}) = \sum_{k = 0}^{+\infty} \frac{(-u^{\nu})^k}{\Gamma(k\nu + 1)}\,,
\quad 0<\nu\le 1,
\end{equation}
that is known to generalize the exponential function 
$\exp(-u)$ to which it reduces just  for $\nu=1$. For details on this transcendental function the reader is referred to the 2014 treatise 
by Gorenflo, Kilbas, Mainardi and Rogosin \cite{GKMR_BOOK14}.
For applications of the Mittag-Leffler  function in linear viscoelasticity based on fractional calculus, we may refer e.g. to Mainardi (1997) \cite{Mainardi_CISM97}, to his 2010 book    \cite{Mainardi_BOOK10}
and  to Mainardi and  Spada (2011) \cite{Mainardi-Spada_2011}.
We recall that in  our numerical calculations we always chose $J_0 = q = 1$, although these  parameters are kept in the expressions, for the sake of  generality. 

We note that the limiting case $\nu =0$ requires special attention because in this case the Mittag-Leffler  function is not defined. However, in this case, by summing according to Ces\`aro the undefined series of the corresponding limit of the 
Mittag-Leffler function, known as Grandi's series
\footnote{
This series is a particular realization of the so called Dirichlet $\eta$
function \cite{NIST}. The latter is part of a broad class of function series, known as
Dirichlet series, more known in rheology as Prony series, that have recently found new physical applications in the so-called Bessel models, see e.g. 
\cite{Giusti-Mainardi_2016,Colombaro-Giusti-Mainardi_2017,Giusti_2017}.
}
\begin{equation}
\label{Grandi's series}
 \sum_{n=0}^\infty (-1)^n =
1-1+1-1+\cdots = \frac{1}{2}\,,
\end{equation}
we get
\begin{equation}
\psi_0 (t) = \frac{t}{2}\,.
\end{equation}
This regularized result  corresponds to the linear creep law for a Maxwell body.
As a consequence, our generalized Becker model is 
{effectively}
defined for $0\le \nu \le 1$  ranging from the Maxwell body
 at $\nu=0$ to the Becker body at 
 $\nu=1$.
 
 In Fig. 3 we show the creep function  
 $\psi_\nu(t)$ in a linear scale $0\le t\le 10$ for the particular values of $\nu=0, 0.25, 0.50, 0.75, 1$,  
from where we can note the tendency to the Maxwell creep law as
$ \nu \to 0^+$.

\begin{figure}[h!]
\begin{center}
\includegraphics[scale=0.65]
{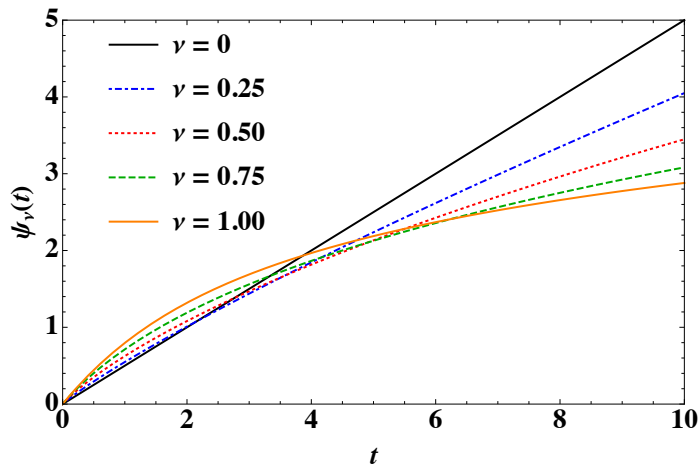}
 \end{center}
 \vskip-0.5truecm
\caption{The creep function 
 for some values of $\nu \in [0,1]$.}
\end{figure}

  \section{{The generalized Becker model: the rate of creep and the spectra}}

For the reader's  convenience  we consider the time derivative of the creep function simply referred to as
the rate of creep  and we show in Fig. 4 the corresponding plots  for different values of the parameter $\nu \in [0,1]$.
Indeed in experimental papers we often find such curves that are observed  to be  decreasing ones, mainly  in the so-called primary stage of creep.    
 \begin{figure}[h!]
\begin{center}
\includegraphics[scale=0.55]
{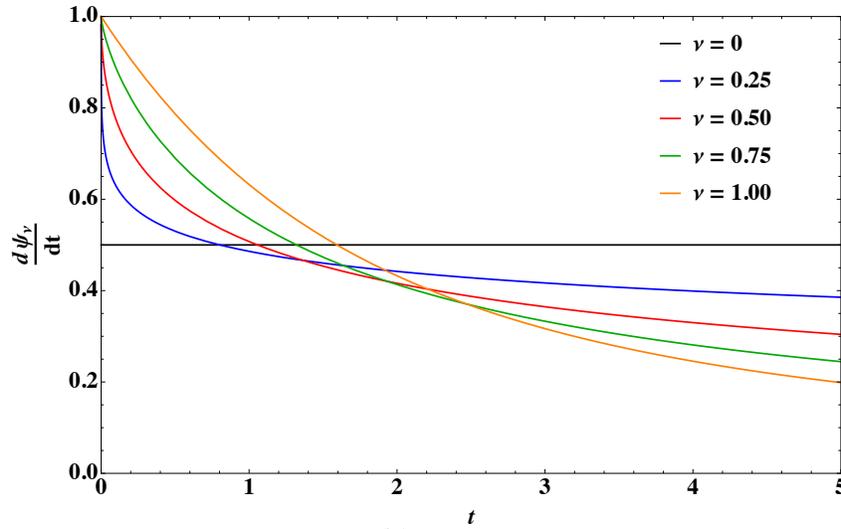}
\end{center}
\vskip-0.5truecm
\caption{The rate of creep  ${\ds \frac{d\psi_\nu}{dt}} (t)$  for different values of $\nu \in [0,1]$.}
\end{figure} 

 {We write the rate of creep obtained from 
\eqref{Psi-nu}, \eqref{Ein-nu} along with their     
  asymptotic representations that are  derived from the asymptotic formulas of the Mittag-Leffler function as reported e.g. in the treatise by Gorenflo et al. (2014) \cite{GKMR_BOOK14}
in the range $0<\nu\le1$}. 

  \begin{equation}
\frac{d\psi_\nu}{dt} (t)=  
 \Gamma(1+\nu)\, \frac{1 - E_{\nu}(-t^{\nu})}{t^{\nu}}
\sim
\left\{
\begin{array}{ll}
{\ds 1- \frac{\Gamma(1+\nu)} {\Gamma(1+2\nu)}\, t^\nu } \,, 
& t \to 0^+\,, \\ \\
{\ds \frac{\Gamma(1+\nu)}{t^\nu}}\,,  & t \to +\infty\,.
\end{array}
\right.
\label{rate-of-creep-nu}
\end{equation}

{We point out  that the rate of creep for our generalized Becker model is still  a CM function like for the standard Becker model in  view of the same theorem we used earlier for proving the CM property. 
 Indeed, in view of  \eqref{rate-of-creep-nu},
the rate of creep 
can be written as
\begin{equation}
\frac{d \psi_\nu}{dt}(t) = 
\Gamma(1+\nu) \left[
\frac{1}{t^\nu} - \frac{E_\nu (-t^\nu)}{t^\nu}\right]   \in[0,1] \quad t\ge0\,,
\quad 0< \nu<1\,.
 \label{rate-of-creep}
\end{equation}
Here we have used the CM property of the Mittag-Leffler function
$E_\nu(-t^\nu)$ for $0<\nu\le 1$, see e.g. \cite{GKMR_BOOK14}.
This statement appears justified  because our model allows a continuous transition between the Maxwell and Becker laws for which the corresponding rates of creep turn out to be CM functions.  Furthermore, it is confirmed  by the non-negativity of  spectra evaluated numerically  by using   
 the Titchmarsh formula \eqref{K(r)} and 
 Eq. \eqref{H(tau)}. Indeed, for $0<\nu<1$, the Laplace transform of the rate of creep  is not known in analytic form  so that it can be obtained  integrating term by term the series representation of 
 the original function. 
This has been carried out  by using the 
{$ MATHEMATICA^{\textregistered}$} tool box.}

{We show in Fig. 5 the spectra in frequency and in time of the rate of the creep for some cases in the range $0<\nu \le 1 $ that turn out to be {effectively}
non-negative (with  a semi-infinite support $[0, +\infty)$ except in  the Becker case $\nu=1$). }
\begin{figure}[h!!]
    \centering
    \includegraphics[width=6.5cm]{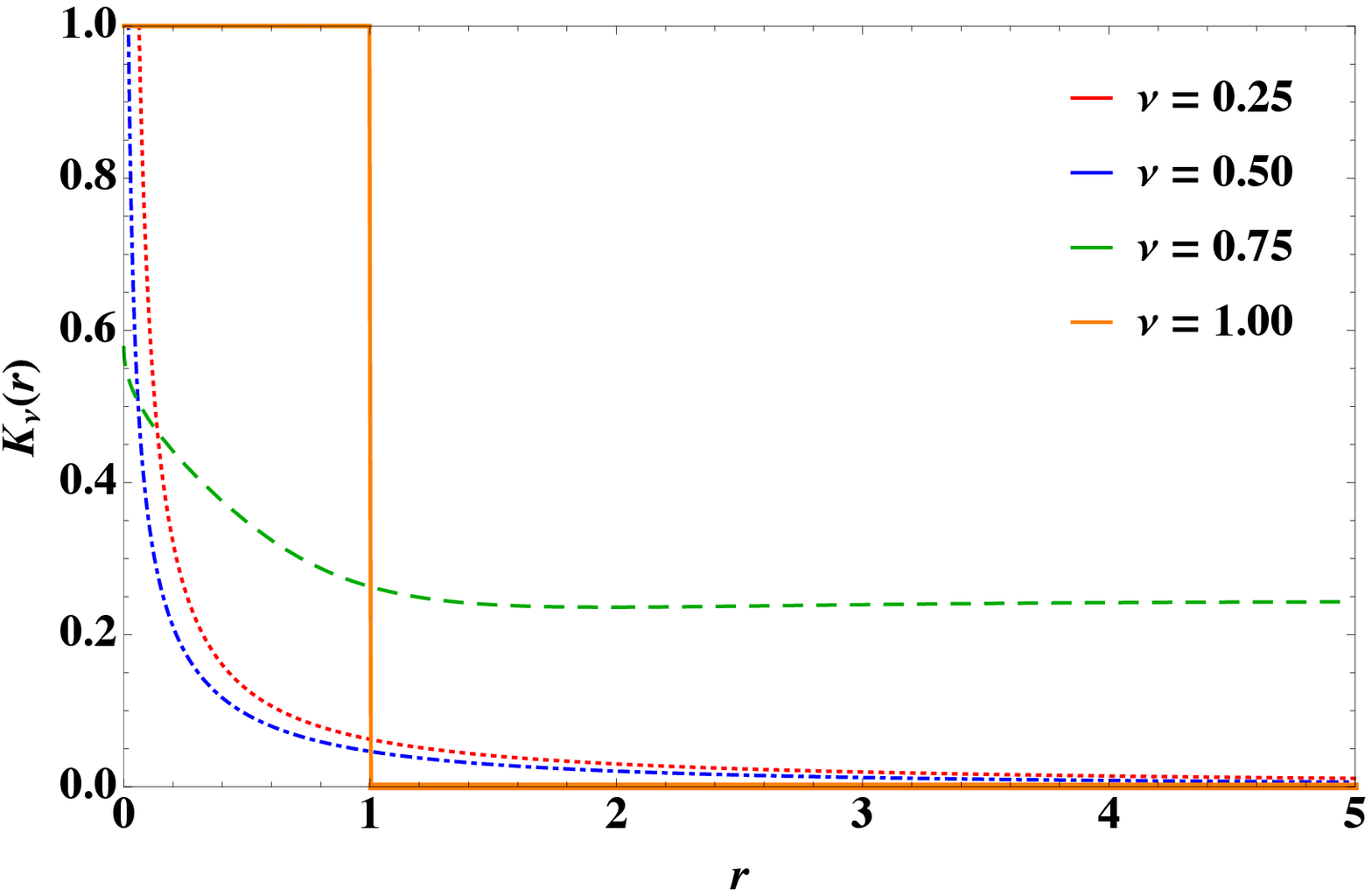}%
    $\quad$
    \includegraphics[width=6.5cm]{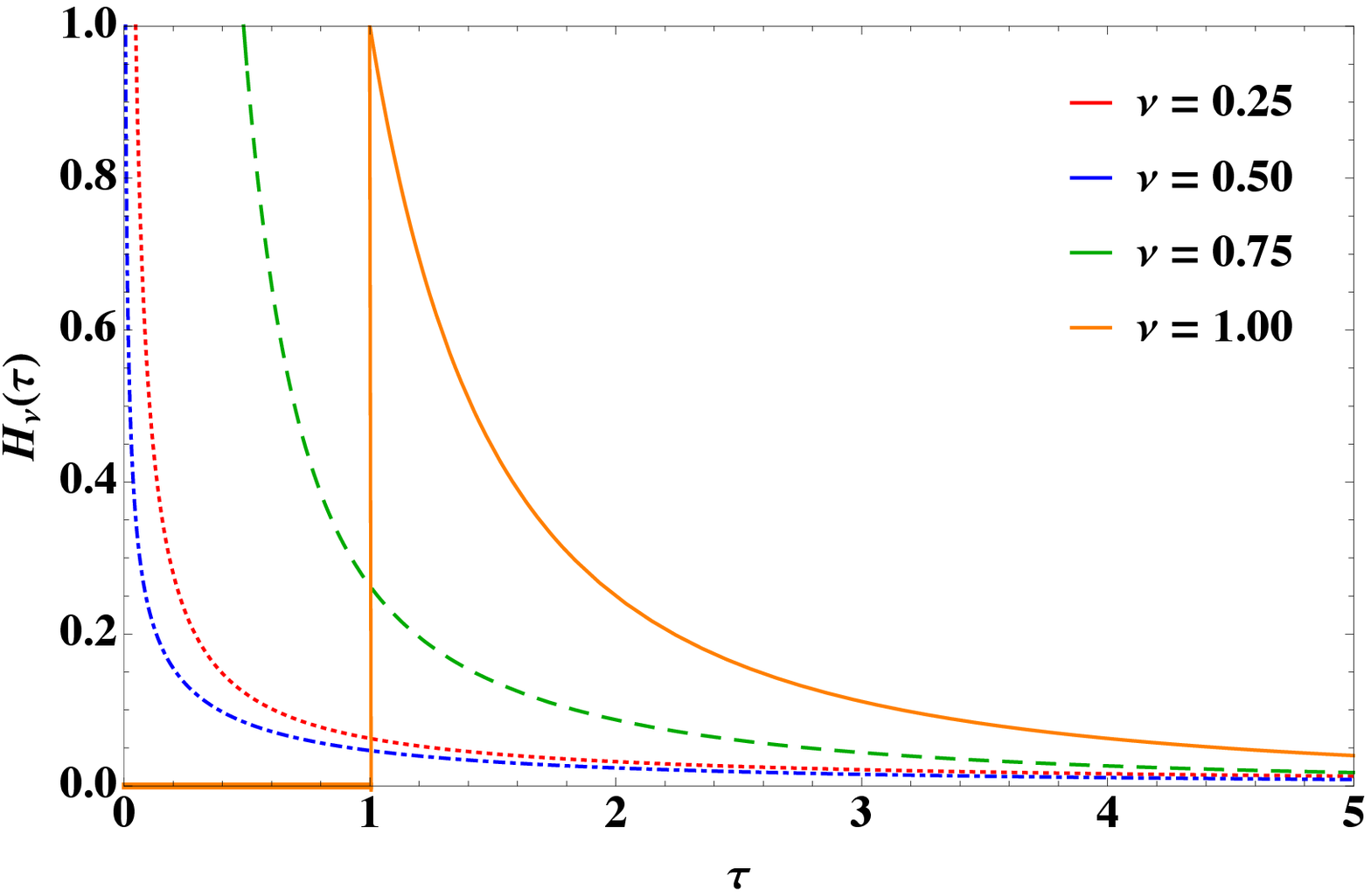} %
    \caption{The spectra for the generalized Becker model  for $\nu=0.25, 0.50, 0.75$ compared with those of the Becker model $\nu=1$: left in frequency $K_\nu(r)$; right in time $H_\nu(\tau)$.}%
\label{H-K TOTALI-spectra}%
\end{figure}



  
  \newpage

\section{The generalized Becker model: {the relaxation function}}

For our generalized Becker {model} we now introduce the  relaxation  modulus:
\begin{equation}
G_{\nu}(t) = G_0\, \phi_\nu(t)
\end{equation}
in terms of the  relaxation function $\phi_{\nu}(t)$ which in turn is  related to the creep function  
by the following Volterra integral equation of the second kind  
\begin{equation}
\phi_{\nu}(t) = 1 - q\, \int_0^t 
\frac{d\psi_{\nu}}{dt'}\, 
\phi_{\nu}(t - t')\,  dt'\,.
\end{equation}
This equation has been solved numerically by using the method already adopted in a recent paper by Garra et al. (2017) \cite{GMS_2017}.
We point out that only in the limiting case $\nu=0$ we get the analytic solution 
\begin{equation}
\phi_0 (t) =   \exp(-q\,t/2)\,,
\end{equation}
 corresponding to the relaxation function of the Maxwell body.
 In  Fig. 6  we show the relaxation function 
 $\phi_\nu(t)$ 
 in a 
 linear scale $0\le t\le 10$ for the particular values of $\nu=0, 0.25, 0.50, 0.75, 1$,
 taking as usual $q=1$. 

 \begin{figure}[h!]
\begin{center}
\includegraphics[scale=0.65]
{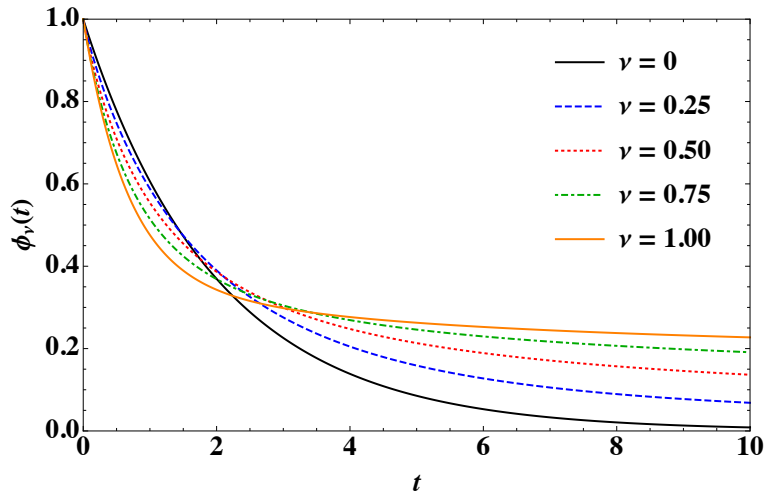}
\end{center}
\vskip-0.5truecm
\caption{The relaxation function $\phi_\nu (t)$   for different values of $\nu \in [0,1]$.}
\end{figure}

\section{The generalized Becker model: the specific dissipation function}
  
  We now consider  the so called \textit{specific dissipation} or {\it internal friction} or {\it loss tangent} related to the dissipation of energy for sinusoidal excitations in stress or strain.
  Referring again to the book by Mainardi (2010) \cite{Mainardi_BOOK10} we use the notation
  \begin{equation}
  \label{Q-def} 
  Q^{-1}(\omega):= 
  \frac{1}{2\pi} \frac{\Delta E}{E_{s}}\,,
  \end{equation}
   most common in Geophysics as a function of a non dimensional angular frequency $\omega$ related to the sinusoidal excitations, where
   $\Delta E$ is the amount of energy dissipated  in one cycle
   and $E_{s}$ is the peak energy stored  during the cycle.
  {For more details see the survey paper by Knopoff in 1964 \cite{Knopoff_RevGeo1964}.}
  
  The final formula, see \cite{Mainardi_BOOK10}, is provided in terms of the complex compliance   related to the Laplace transform of the strain compliance $J(t)$ and reads 
  \begin{equation}
{Q}^{-1}(\omega) =  
\left.
\frac{\Im\{s \widetilde J(s)\} }
{\Re\{s\widetilde J(s)\}}
\right|_{s = \pm i\omega}\,,
\end{equation}
where the positive result must be taken for $\omega>0$.
   As  a consequence, for our generalized Becker model depending on the parameter $\nu \in [0,1]$, we get
   \begin{equation}
{Q}^{-1}_{\nu} (\omega) = 
\left.
\frac{
\Im[1 + q\,s\widetilde\psi_{\nu}(s)}
{\Re[1 + q\, s\widetilde\psi_{\nu}(s)]}
\right|_{s = \pm i\omega}\,.
\end{equation}  
Analytic expressions are expected to be only available in the limiting cases
$\nu=0$ (Maxwell model) and $\nu=1$ (Becker model).

After regularization of Grandi's series,
for the Maxwell model we get 
$$\psi_0(t) = \text{Ein}_0(t)
= \frac{t}{2}\,, \quad
\frac{d\psi_0}{dt} = \frac{1}{2}\,,$$
so
$$s \tilde \psi_0(s) = \mathcal{L}\left\{
\frac{d\psi_0}{dt}\right \} = 
\mathcal{L}\left\{\frac{1}{2}\right \} = \frac{1}{2s}$$
Hence
\begin{equation}
{Q}^{-1}_{0} (\omega) = \frac{\Im\left[1 + \frac{q}{2s}\right]}{\Re\left[1 + \frac{q}{2s}\right]}\bigg|_{s = \pm i\omega} = \frac{\Im\left[1 \pm \frac{q}{2i\omega}\right]}{\Re\left[1 \pm \frac {q}{2i\omega}\right]} = \Im\left[1 \mp \frac{iq}{2\omega}\right] = \frac{q}{2\omega}\,,
\end{equation}
{which constitutes} a well known result.

For the Becker model we get 
$$\psi_1(t) = \text{Ein}(t)\,, \quad \frac{d\psi_1}{dt} = \frac{1-\e^{-t}}{t}\,,$$
so
$$s \tilde \psi_1(s) = 
\mathcal{L}\left\{\frac{d\psi_1}{dt}\right \} =
 \mathcal{L}\left\{\frac{1 -\ e^{-t}}{t}\right \} = 
 \ln\left(1 + \frac{1}{s}\right)\,.$$
The specific dissipation is then given after some calculations of complex analysis 

\begin{equation}
{Q}^{-1}_1(\omega) = {\ds
\frac{2q\, \arctan\left(\sqrt{1+ \omega^2} - \omega\right)}
{1 +q\, \ln {\ds \frac{\sqrt{1 + \omega^2}} {\omega} } } }
= {\ds \frac{2q\, 
 \arctan\left(\sqrt{1+\omega^2 } - \omega\right)}
 {1 + q
 \left( \frac{1}{2}\ln(1 + \omega^2) - \ln(\omega)\right)}}\,,
\end{equation}
 which can be simplified as  follows
\begin{equation}
Q^{-1}_1(\omega) = 
{\ds \frac{\arctan\left(1\over\omega\right)}
{
{\ds 
\frac{1}{q} +
 \frac{1}{2}\ln\left(\frac{1 + \omega^2}{\omega^2}\right)}
 }}\,.
\end{equation}
We note that this expression coincides with Eq. (8) found by Strck and  Mainardi (1982)
\cite{Strick-Mainardi_1982}, where the authors have assumed $\tau= 10^{-10} s$
and $1/q= 57.812$ in order to have a dissipation function compatible with some experimental data in Seismology.
 
We also note that whereas for the Maxwell model the specific dissipation  decreases from infinity to zero in the range 
$0<\omega<\infty$, in the Becker model the specific dissipation increases from zero at $\omega =0$ to a certain value at an intermediate frequency and then decreases  to zero as 
$\omega \to \infty$. 

In Fig. 7 and in Fig. 8 
we show the specific dissipation function 
 $Q^{-1}_\nu (\omega)$ by adopting linear and logarithmic scales respectively,  
 for the particular values of $\nu=0, 0.25, 0.50, 0.75, 1$,
 taking as usual $q=1$. 
 
 From our plots we recognize that in the intermediate cases $0<\nu<1$ the specific dissipation assumes a finite value at 
$\omega=0$ decreasing with increasing $\nu$.  Then for $\nu \ge 0.75$ the function increases up to a  maximum whereas for 
$\nu = 0, 0.25, 0.50$ it is   {always} decreasing. The transition value of $\nu$ in the interval 
$0.50< \nu<0.75 $ for the existence of such a maximum cannot be analytically determined.   
In any case,  {for $\nu>0$},  it is possible to find a frequency range
 where the dissipation factor is almost constant 
  by taking a suitable factor $q$, as it was required by Becker in his 1925 paper.

\newpage 
  
  \begin{figure}[h!]
\begin{center}
\includegraphics[scale=0.65]
{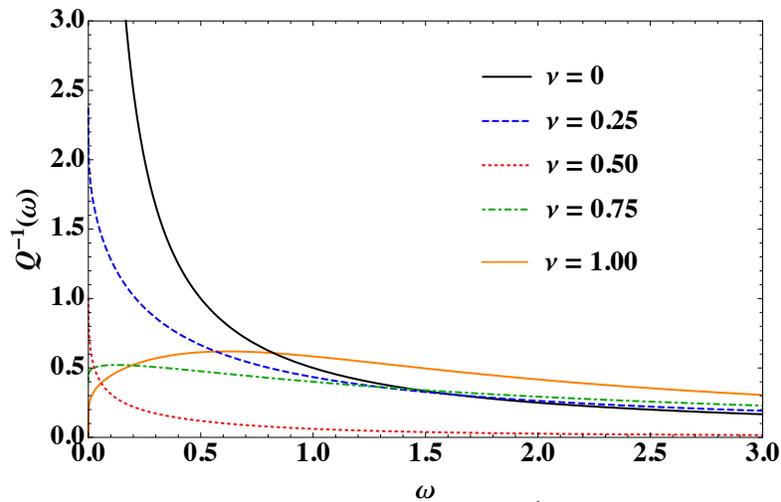}
\end{center}
\vskip-0.5truecm
\caption{Specific dissipation  function 
$Q^{-1}_\nu(\omega)$ for different values of $\nu  \in [0,1]$,  
by adopting linear scales.}
\end{figure} 
  
  \vskip 1truecm
  
  \begin{figure}[h!]
\begin{center}
\includegraphics[scale=0.8]
{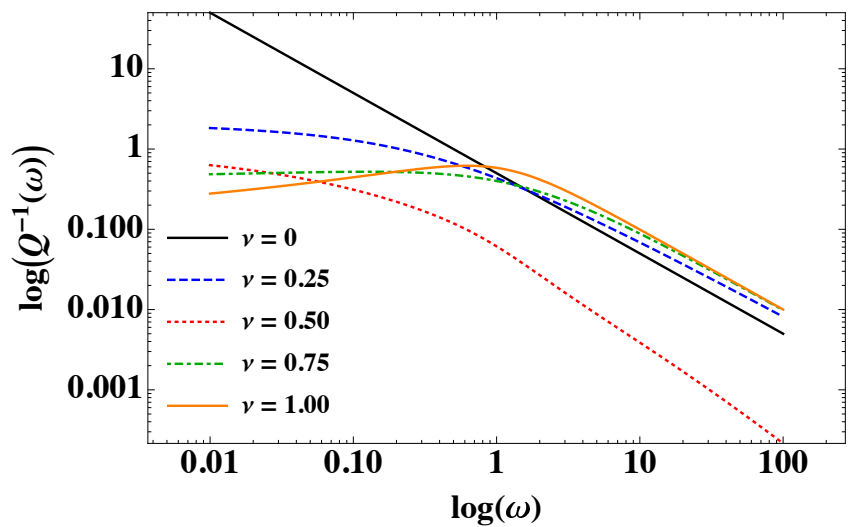}
\end{center}
\caption{Specific dissipation  function 
$Q^{-1}_\nu(\omega)$ 
for different values of 
$\nu  \in [0,1]$,  by adopting logarithmic  scales.}
\end{figure}

\newpage

\section{Conclusions}

 We have presented  a new rheological  model starting from the creep law of the so-called Becker body.
 Indeed we have generalized the Becker creep law by introducing  the Mittag-Leffler  function  of order $\nu \in  (0,1)$
  that in the limit of vanishing $\nu$  allows us to  recover the linear creep law of the Maxwell body.
 We recall that a different transition in creep from   a linear behavior
 typical of the Maxwell body to a logarithmic behavior typical of the Lomnitz model has been investigated by Mainardi and Spada (2012b) 
 \cite{Mainardi-Spada_2012b}
        for the Strick-Jeffreys-Lomnitz model of linear viscoelastity.

Then,   based on the hereditary theory of linear viscoelasticity,
for our generalized Becker model  
     we have also  {approximated 
    the corresponding relaxation function by solving numerically a Volterra integral equation of the second kind.
     The problem  of the evaluation of the corresponding spectral distributions has been left to  a future paper.}
   Furthermore,  we have provided a full characterization of the new model  adding to the  creep and  relaxation functions the so-called specific dissipation $Q^{-1}$  function versus frequency of relevance in Geophysics.   
   
Illuminating    plots of the characteristic functions (creep, relaxation, specific dissipation) have been presented for the readers' convenience.
 This similarity  (apart from a suitable scaling factor) 
{leads us to believe} 
 that  our generalized model can hopefully be assumed by experimentalists in rheology  to fit some curves of these characteristic  functions in their experiments.   
We do hope that the results obtained  in this paper may be useful for fitting  experimental data  in rheology of real materials that exhibit 
   responses in creep, relaxation  {and energy dissipation} varying between the  Maxwell and Becker bodies.
We are thus  confident to have found a suitable application of the Mittag-Leffler function  in linear viscoelasticity 
 without    involving the    possible constitutive stress-strain equations of fractional order.

A systematic comparison between  {the theoretical responses}  predicted by the generalized Becker 
model and experimental results is out of the scope of present paper. However, we have noted that the Becker creep law has {already} found some application in the rheology of the Earth mantle and of the primary creep of ice. In particular, for ice
it was  found that the fit of experimental data with creep functions containing exponential integrals is not fully satisfactory. Hence, the generalization of the Becker law by the introduction of an extra parameter $\nu$  via the Mittag-Leffler function  that we have accomplished here could potentially help to 
improve the agreement with experimental data. This will be the subject of a follow up paper.
{Furthermore, because the Mittag-Leffler function enters in any  creep and relaxation function of the fractional viscoelastic models, see e.g. 
{Caputo and Mainardi (1971) \cite{Caputo-Mainardi_RNC71}},   Gl\"ockle and  Nonnenmacher (1991) \cite{Glockle-Nonnenmacher_1991},
Metzler et al. (1995) \cite{Metzler-et-al_1995},  Mainardi (2010) \cite{Mainardi_BOOK10} 
{and  Sasso et al. (2011) \cite{Sasso-et al_2011}}  for more references,
we also expect that the constitutive law of our generalized Becker model could be based on a differential stress-strain relation of fractional order. } 

We close this section with  this relevant statement again: 
for $0< \nu<1$ the corresponding non dimensional functions $\psi_\nu(t)$ and $\phi_\nu(t)$ keep the property to be Bernstein and CM functions as  for the Maxwell ($\nu=0$) and Becker ($\nu=1$) bodies, which implies that our generalized Becker model
is in agreement with  basic physical principles of linear viscoelasticity.}


 \section*{{Societal value
of the presented research results}}
  {This study could lead to better knowledge of the mechanical properties of some materials, with possible applications to engineering and industry.
  Indeed rheology is relevant in the mechanics  of time-dependent materials and this model, depending on a parameter, could be assumed by experimentalists to better fit their curves on creep, relaxation and energy disspation at most in the primary stage of deformation where linear viscoelasticity is dominant.}

  \section*{Acknowledgments}
	The work of F. M. has been carried out in the framework of the activities of the National Group of Mathematical Physics (GNFM, INdAM).
The work of G. S. has been carried out in the framework of the activities of the Department of Pure and Applied Sciences (DiSPeA) of the Urbino University  "Carlo Bo".	
		The authors would like to thank  Alexander Apelblat,  Andrea Giusti, Andrzej Hanyga 
		and Nanna  Karlsson  for valuable comments, advice  and discussions.
{Furthermore, the authors are grateful  to the anonymous referees for careful reading of our paper and for making several important suggestions which improve the presentation.}  

\appendix
\section{Essentials of linear viscoelasticity}

  We recall that in the linear   theory of viscoelasticity, based on the hereditary theory by Volterra,  
a viscoelastic body  is characterized by two distinct but interrelated material functions,
causal in time (i.e. vanishing for $t<0$):
the creep compliance $J(t)$ (the strain response to a unit step of stress)
and the relaxation modulus  $G(t)$ (the stress response to a unit step of strain).
For more details, see e.g.  
Christensen (1982) \cite{Christensen_BOOK82},
Pipkin (1986)  \cite{Pipkin_BOOK86},
Tschoegl (1989) \cite{Tschoegl_BOOK1989},
 Tschoegl (1997) \cite{Tschoegl_1997} 
 and 
Mainardi (2010) \cite{Mainardi_BOOK10}.

By taking $J(0^+)=J_0 >0$ so that $G(0^+)= G_0 =1/J_0$,
the body is assumed to exhibit a non vanishing  instantaneous response both in the  creep  and in the relaxation tests.
 As a consequence, we find it convenient to introduce two  dimensionless quantities $\psi(t)$ and $\phi(t)$  as follows 
\begin{equation}
\label{J-G-definitions}
J(t)= J_0[1+q\, \psi(t)]\,, \quad G(t) = G_0\, \phi(t)\,,
\end{equation}  
where $\psi(t)$ is a non-negative  increasing function with $\psi(0) =0$ and  
$\phi(t)$ is a non-negative decreasing function with $\phi(0)=1$. 
We have assumed, without loss of generality $\tau_0=1$, but we have kept the non-dimensional quantity $q$  for a suitable scaling of the strain, according to convenience in experimental  rheology. 
At this stage, viscoelastic bodies may be distinguished in solid-like 
and fluid-like  whether $J(+\infty)$  is finite or infinite so that 
$G(+\infty)= 1/J(+\infty)$ is non zero or zero, correspondingly.  

   As pointed out  {in most treatises on linear viscoelastity}, e.g. in  
\cite{Pipkin_BOOK86}, \cite{Tschoegl_BOOK1989},    
   \cite{Mainardi_BOOK10},
    the relaxation modulus $G(t)$      
    can be derived from the corresponding creep compliance $J(t)$  through the Volterra integral equation of the second kind
    \begin{equation}
    G(t)= \frac{1}{J_0}-\frac{1}{J_0}\,
    \int_0^t \!\frac{dJ}{dt'}\, G(t-t')\, dt'\, ;
    \label{integral-equation}
    \end{equation}
    then, as a consequence of  
    Eq. \eqref{J-G-definitions}, 
    the non-dimensional relaxation function $\phi(t)$ 
    obeys the Volterra integral equation
    \begin{equation}
    \label{integraleqphi}
    \phi(t)=1-q\, \int_0^t\frac{d\psi}{dt'}\, \phi(t-t')\, dt'\,.
    \end{equation}

  In linear viscoelasticity,
 it is quite common  
   to require the existence of positive retardation and relaxation spectra for the material functions $J(t)$ and $G(t)$,
 as pointed out  by  Gross in his 1953 monograph on the mathematical structure of the theories of viscoelasticity \cite{Gross_BOOK53}.
  This implies,  as  formerly proved in 1973 by Molinari \cite{Molinari_1973}  and revisited  in 2005 by Hanyga \cite{Hanyga_2005}, see also  Mainardi's book    \cite{Mainardi_BOOK10},
that $J(t)$ and $G(t)$  {and consequently the functions $\psi(t)$ and $\phi(t)$}
turn out to be Bernstein and Completely Monotonic (CM) functions, respectively.       

 {Here we  recall that a CM function $f(t)$  is a non negative, infinitely derivable function with derivatives alternating in sign for $t>0$ like $\exp(-t)$, whereas a Bernstein function is a non negative function whose derivative is CM, like $1-\exp(-t)$.}
{Then, a necessary and sufficient condition to be a CM function is provided by the Bernstein theorem according to which $f(t)$ is the Laplace transform of a non-negative real function.}
For more details on  these mathematical properties  the interested reader is referred to  the excellent monograph by Schilling et al. \cite{sch}.

For the rate of creep, we  write
\begin{equation}
\label{CM-spectra}
\frac{d\psi}{dt} (t)= \int_0^\infty \!\!
\e^{-rt}\, K(r)\, dr = 
\int_0^\infty \!\! \e^{-t/\tau}\, H(\tau)\, d\tau \,, 
\end{equation}
where  $ K(r)$ and  $H(\tau)$ are  the required spectra
in frequency ($r$) and in time ($\tau=1/r$), respectively.

  The frequency spectrum can be determined from the Laplace transform of the rate of creep by the Titchmarsh formula that reads in an obvious notation,   if  $\psi(0^+)=0$, 
\begin{equation}
\label{K(r)}
K(r) = \pm \frac{1}{\pi}\,
\left.
\Im[s\widetilde\psi (s)]
\right|_{s =r \e^{\mp i\pi}}\,.
\end{equation}
This a consequence of the fact that the Laplace transform of the rate of creep is the iterated Laplace transform of the frequency spectrum, that is the Stieltjes transform of $K(r)$ and henceforth the Titchmarsh formula provides   the inversion of the  Stieltjes transform, 
{see e.g. Widder (1946) \cite{Widder_LT1946}.}
As a consequence, the time spectrum can  be determined using the transformation $\tau= 1/r$, so that
\begin{equation}
\label{H(tau)}
H(\tau) =
\frac{K(1/\tau)}{\tau^2}\,.
\end{equation}




\begin{thebibliography}{99}
  
 
      \bibitem{Becker_1925}
R. Becker,  
Elastische Nachwirkung und Plastizit\"at,
{\it Zeit. Phys.} {\bf 33} (1925), 185--213.

\bibitem{Becker-Doring_BOOK39}
 R. Becker  and  W. Doring,  
  {\it Ferromagnetismus}, Springer-Verlag, Berlin (1939).
   
 \bibitem{Caputo-Mainardi_RNC71}
 M. Caputo and F. Mainardi,
Linear models of dissipation in anelastic solids,
{\it Rivista del Nuovo Cimento (Ser.II)} {\bf  1} (1971),  161--198. 
   
      \bibitem{Christensen_BOOK82}
R.M.  Christensen, 
 {\it Theory of Viscoelasticity}, 2-nd edition,
Academic Press, New York (1982).  [First edition (1972)]

\bibitem{Colombaro-Giusti-Mainardi_2017}
I. Colombaro, A. Giusti and F. Mainardi, A class of linear viscoelastic models based on Bessel functions,
{\it  Meccanica} (2017) {\bf 52} No 4-5 (2017), 825--832; 
DOI: 10.1007/s11012-016-0456-5
[E-print: https://arxiv.org/pdf/1602.04664.pdf]



 \bibitem{GMS_2017}
R. Garra, F. Mainardi and G. Spada,
A generalization of the Lomnitz logarithmic creep law via fractional calculus,
{\it Chaos, Solitons and Fractals}
Printed on line 30 March (2017);
DOI: 10.1016/j.chaos.2017.03.032
[E-print: https://arxiv.org/pdf/1701.03068.pdf]

\bibitem{Giusti_2017}
A. Giusti, 
On infinite order differential operators in fractional viscoelasticity,
{\it  Fract. Calc. Appl. Anal.} {\bf  20} No 4 (2017), 854--867;
DOI: 10.1515/fca-2017-0045
[E-print https://arxiv.org/pdf/1701.06350.pd]


 \bibitem{Giusti-Mainardi_2016}
 A. Giusti and F. Mainardi, 
 A dynamic viscoelastic analogy for fluid-filled elastic tubes.
 {\it Meccanica} {\bf 51} No 10 (2016), 2321-2330; 
DOI: 10.1007/s11012-016-0376-4
[E-print: https://arxiv.org/pdf/1505.06694.pdf]
  
\bibitem{Glockle-Nonnenmacher_1991}
W.G. Gl\"ockle and T. Nonnenmacher,
Fractional integral operators and Fox  functions in the theory of viscoelasticity,
{\it Macromolecules} {\bf 24} (1991),  6426--6434.

    \bibitem{GKMR_BOOK14} 
      R. Gorenflo, A.A. Kilbas, F. Mainardi and  S.V. Rogosin, 
      \emph{Mittag-Leffler Functions, Related Topics and Applications}, 
      Springer, Berlin (2014). [2-nd edition in preparation]



\bibitem{Gross_BOOK53}
 B. Gross, 
{\it Mathematical Structure of the Theories of Viscoelasticity},
Hermann \& C., Paris (1953).      
      
    
           	 
\bibitem{Hanyga_2005}
A. Hanyga, 
Viscous dissipation and completely monotonic relaxation moduli,
{\it Rheologica Acta} {\bf 44} (2005), 614--621.

\bibitem{Hanyga_PAGEOPH2014}
A. Hanyga, 
Attenuation and shock waves in linear hereditary viscoelastic media: Strick-Mainardi, Jeffreys-Lomnitz-Strick and Andrade creep compliances,
{\it Pure Appl. Geophys.} {\bf 171} No 9 (2014), 2097--2109.

\bibitem{Holenstein-et-al_JBE1980}
{R. Holenstein, P. Nieder and M. Anliker,
A viscoelastic model for use in predicting arterial pulse waves,
{\it J Biomech. Eng.} {\bf 102} (1980), 318--325.} 

\bibitem{Jellinek-Brill_1956}
H.H.G. Jellinek and R. Brill,
Viscoelastic properties of ice,
{\it J. Appl. Phys.} {\bf 27} No 10 (1956), 1198--1209.

\bibitem{Knopoff_RevGeo1964}
{L. Knopoff, Q,
{\it Rev. Geophys.} {\bf 2} No 4 (1964), 625--660.}


\bibitem{Lubliner-Panoskaltsis_1992}
   J. Lubliner and  V.P. Panoskaltsis,
   The modified Kuhn model of linear viscoelasticty,
 {\it   Int. J. Solid Structures} 
 {\bf 29} No 24 (1992), 3099-3112.  
    
  

      	 
      	 
      
\bibitem{Mainardi_CISM97}
F. Mainardi, 
Fractional calculus, some basic problems in continuum and statistical mechanics,
in :
A. Carpinteri,  and F. Mainardi  (Editors),
{\it Fractals and Fractional Calculus in  Continuum Mechanics},
Springer Verlag, Wien and New York (1997), pp. 291--348.
[E-print  http://arxiv.org/abs/1201.0863]
      
      
\bibitem{Mainardi_BOOK10}
F. Mainardi,  \textit{Fractional Calculus and Waves in Linear Viscoelasticity},
Imperial College Press, London (2010). 
[2-nd edition in preparation]


\bibitem{Mainardi-Spada_2011}
     F. Mainardi and G. Spada,
 Creep, relaxation and viscosity properties for basic fractional models in rheology,
{\it The European Physical Journal, Special Topics} {\bf  193} (2011),  133--160.
[E-print: http://arxiv.org/abs/1110.3400]
  
\bibitem{Mainardi-Spada_2012a}
      F. Mainardi and G. Spada,
        Becker and Lomnitz rheological models: A comparison, 
        in A. D'Amore, L. Grassia and  D. Acierno (Editors), 
AIP (American Institute of Physics)  Conf. Proc. Vol. 1459, pp. 132-135 (2012). 
  Proceedings of  the International Conference TOP (Times of Polymers \& Composites), Ischia, Italy, 10-14 June 2012.
[E-print: http://arxiv.org/abs/1210.5717]

\bibitem{Mainardi-Spada_2012b}
F. Mainardi and G. Spada,
On the viscoelastic characterization of the
Jeffreys--Lomnitz law of creep,
{\it Rheol Acta} {\bf 51}  (2012),  783--791.
[E-print: http://arxiv.org/abs/1112.5543]

 \bibitem{Masina-Mainardi_2018}
E. Masina and F. Mainardi, On  the generalized  Exponential Integral,
Pre-print, Department of Physics, University of Bologna,
{\it to be submitted} (2017).
   
  
\bibitem{Metzler-et-al_1995}
 R. Metzler, W. Schick, H.G.Kilian and T.F. Nonnenmacher,
 Relaxation in filled polymers: a fractional calculus approach,
 {\it J. Chemical Phys.} {\bf 103} No 16 (1995), 7180--7186.
 
 
\bibitem{Molinari_1973}
 A. Molinari,  
Visco\'elasticit\'e lin\'eaire et fonctions compl\`etement monotones,
{\it Journal de M\'ecanique} {\bf 12} (1973),  541--553.

\bibitem{Neubert_AQ1963}
  {H.K.P. Neubert, 
  A simple model representing internal damping in solid materials,
  {\it Aeronautical Quarterly} {\bf 14} No 2 (1963), 187--197.}
      
      \bibitem{NIST}
    {\it  NIST Digital Library of Mathematical Functions}, 
      edited by 
      F.W.J. Olver,
D.W. Lozier,
R. F. Boisvert,
and C.W. Clark,
Cambridge University Press, Cambridge (2010).
   
   
   \bibitem{Orowan_1967}
   E. Orowan, 
   Seismic damping and creep in the mantle, 
   {\it Geophys J. R astr. Soc.} {\bf   14} (1967),  191--218.
   
   
 \bibitem{Pipkin_BOOK86}
 A.C. Pipkin, 
{\it Lectures on Viscoelastic Theory},
2-nd Edition,
Springer Verlag, New York (1986).  [First Edition 1972]

	
  \bibitem{Sasso-et al_2011}
  {M. Sasso, G. Palmieri and D. Amodio,
  Application of fractional derivative models in linear viscoelastic problems,
   {\it Mech. Time-Dependent Materials} {\bf 15} No 4 (2011), 367--387.}
    
            \bibitem{Schelkunoff_1944}
S.A. Schelkunoff,  
Proposed symbols for the modified
cosine and integral exponential integral,
 {\it Quart. Appl. Math.}  {\bf 2} (1944), p. 90.
 
      
       \bibitem{sch} R.L. Schilling, R. Song, Z. Vondracek,
        \emph{Bernstein Functions: Theory and Applications}, 
        De Gruyter (2012).
       
\bibitem{Strick-Mainardi_1982}
E. Strick and F. Mainardi,
  On a general class of constant $Q$ solids,
   {\it Geophys. J.  Roy. Astr. Soc.}
   {\bf 69} (1982),   415--429. 
   
  \bibitem{Tschoegl_BOOK1989} 
N.W. Tschoegl, 
{\it The Phenomenological Theory of Linear Viscoelastic Behavior},
Springer Verlag, Heidelberg (1989).

\bibitem{Tschoegl_1997}
N.W. Tschoegl, 
Time dependence in materials properties: an overview,
{\it  Mechanics of Time-Dependent Materials}
{\bf 1} (1997),  3--31. 


\bibitem{Widder_LT1946}
{D.V. Widder,
{\it The Laplace Transform},  
Princeton University Press, Princeton (1946).}

\end{thebibliography}
        \end{document}